\documentclass[prb,superscriptaddress,showpacs,a4paper,twocolumn,floatfix]{revtex4}

\usepackage{amsmath} 
\usepackage{graphics} 
\usepackage{dcolumn} 
\usepackage{epsfig} 

\begin{document}

\title{Adsorption-Induced Distortion of F$_{16}$CuPc on Cu(111) and Ag(111):
  \\ An X-ray Standing Wave Study}

\author{A. Gerlach} 
\author{F. Schreiber}
\email[corresponding author:]{frank.schreiber@chem.ox.ac.uk}

\affiliation{Physical and Theoretical Chemistry Laboratory, Oxford University,
  South Parks Road, OX1 3QZ, United Kingdom} 

\author{S. Sellner}
\author{H. Dosch} 

\affiliation{Max-Planck-Institut f\"ur Metallforschung, Heisenbergstr. 3,
  70569 Stuttgart, Germany}

\affiliation{Institut f\"ur Theoretische und Angewandte Physik, Universit\"at
  Stuttgart, Pfaffenwaldring 57, 70550 Stuttgart, Germany}

\author{I. A. Vartanyants} 

\affiliation{HASYLAB, DESY, Notkestr. 85, 22607 Hamburg, Germany} 

\author{B. C. C. Cowie}
\author{T.-L. Lee}
\author{J. Zegenhagen}

\affiliation{ESRF, 6 Rue Jules Horowitz, B.P. 220, 38043 Grenoble Cedex 9,
  France}

\date{\today}

\begin{abstract} 
  The adsorption geometry of perfluorinated copper-phthalocyanine molecules
  (F$_{16}$CuPc) on Cu(111) and Ag(111) is studied using X-ray standing waves.
  A detailed, element-specific analysis taking into account non-dipolar
  corrections to the photoelectron yield shows that on both surfaces the
  molecules adsorb in a lying down, but significantly distorted configuration.
  While on copper (silver) the central carbon rings reside $2.61\,$\AA{}
  ($3.25\,$\AA) above the substrate, the outer fluorine atoms are located
  $0.27\,$\AA{} ($0.20\,$\AA) further away from the surface.  This non-planar
  adsorption structure is discussed in terms of the outer carbon atoms in
  F$_{16}$CuPc undergoing a partial rehybridization ($sp^2 \rightarrow sp^3$).
\end{abstract}

\maketitle

\section{Introduction}
\label{sec:intro}

The adsorption of organic molecules on various surfaces has become a subject
of wide interest. With the realization of new organic based semiconductor
devices\cite{forrest_apl01,bao_jacs98} it has been recognized that the first
molecular layer of organic thin films strongly influences their structural and
electronic properties. Hence increasing efforts are being made to improve our
still fragmentary understanding of the complex interaction of aromatic
molecules with metal substrates. A variety of surface sensitive techniques are
being used to explore organic thin films in the monolayer regime.  Low energy
electron diffraction (LEED)\cite{braun_ss01,breitbach_ss02}, photoelectron
diffraction (PED)\cite{schaff_ss96,kang_ss00}, and scanning tunneling
microscopy (STM)\cite{lippel_prl89,chizhov_la00,grand_ss96,boehringer_prb97},
for example, have been successfully employed in this area.

\begin{figure}[htbp]
  \centering
  \includegraphics[width=4cm]{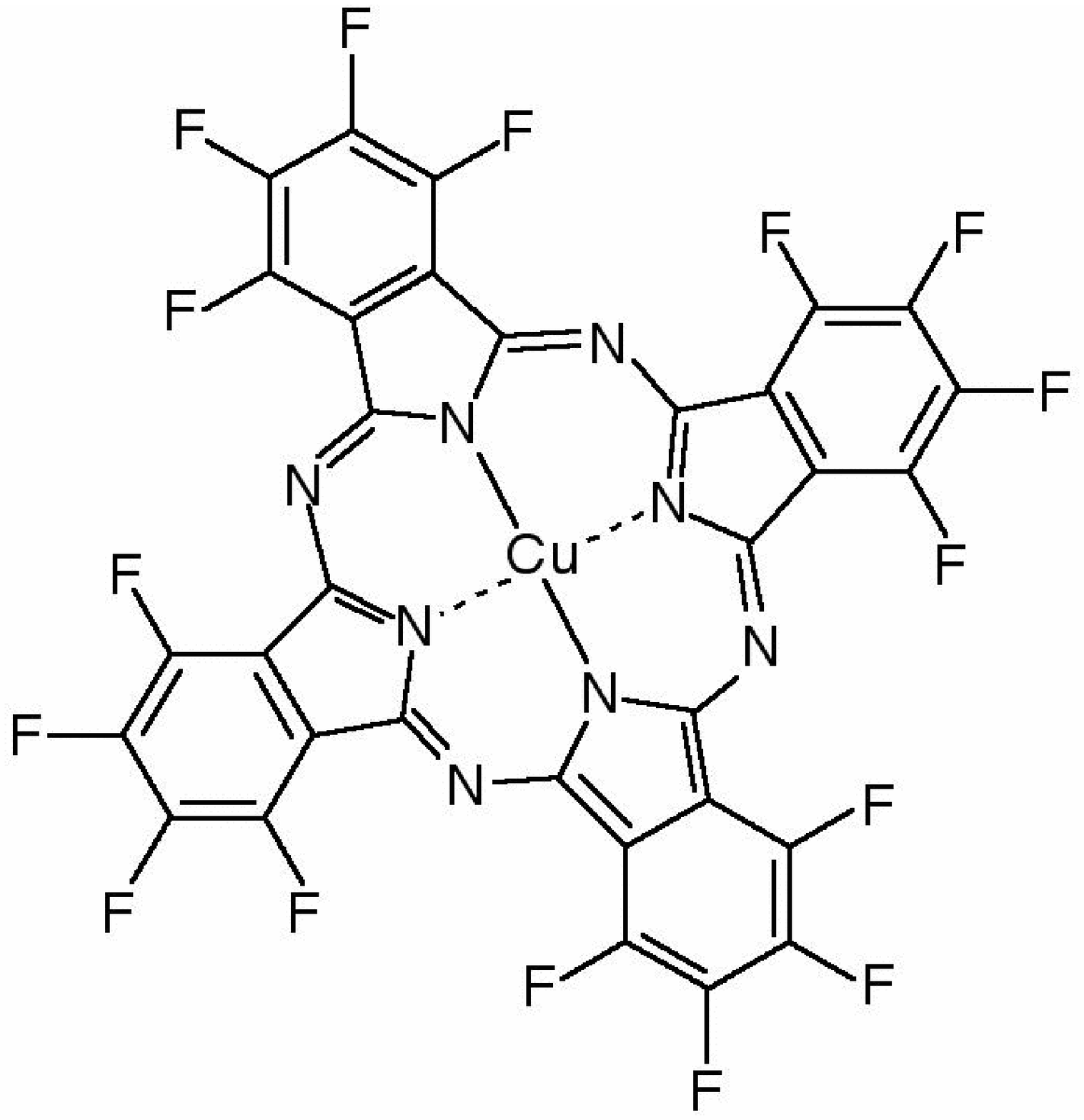}%
  \vspace{1mm}
  \includegraphics[width=6cm]{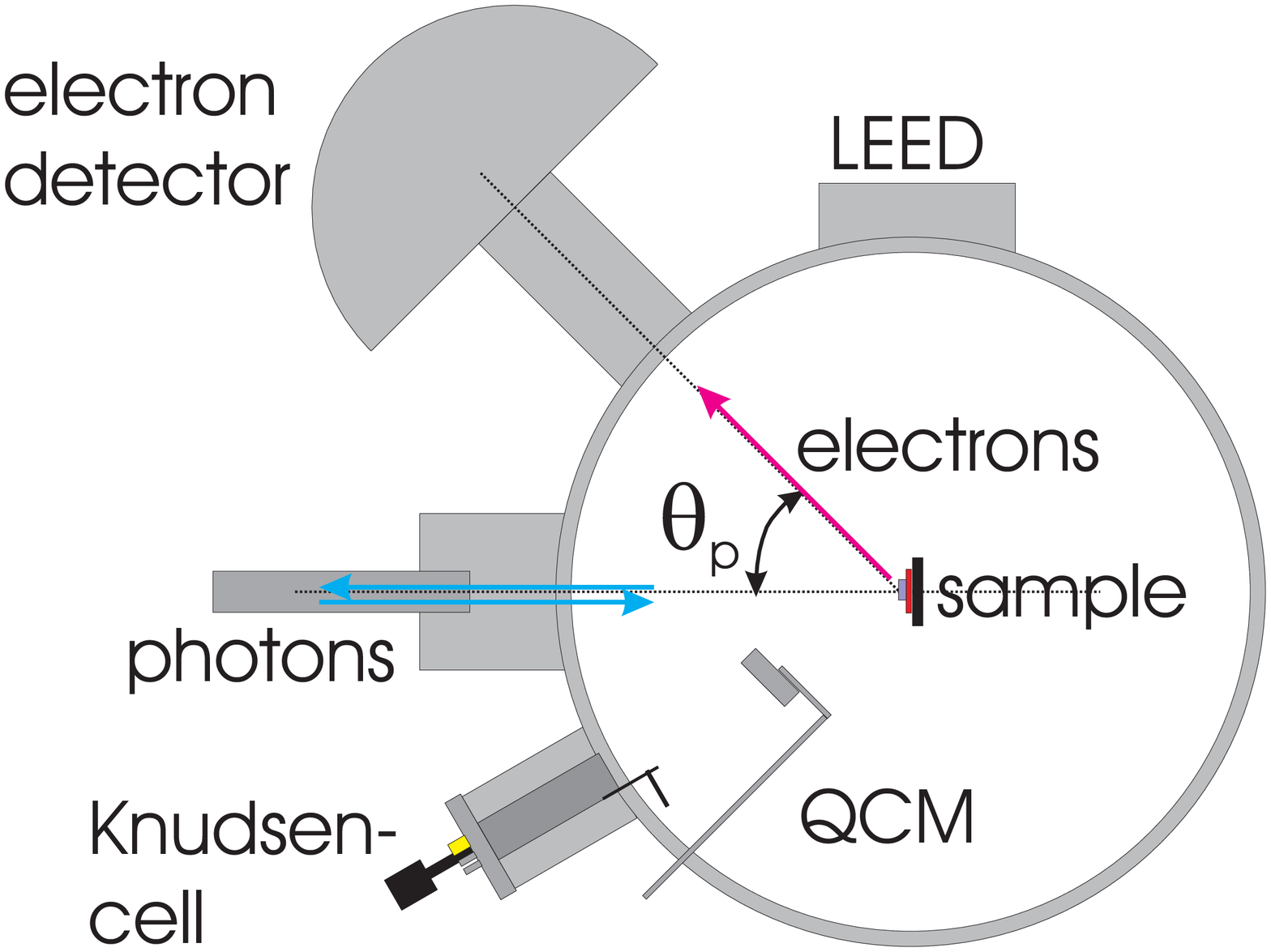} 
  \caption{(Color online) (a) Perfluorinated copper-phthalo\-cyanine
    (F$_{16}$CuPc). (b) Experimental setup at the X-ray standing wave beamline
    ID32 (ESRF).}
  \label{fig:xswsetup}
\end{figure}

When studied in more detail, aromatic molecules exhibit a non-trivial
adsorption behavior, benzene on various substrates being the simplest and
best-studied example.\cite{braun_ss01,breitbach_ss02,schaff_ss96,kang_ss00}
Because of the relatively strong adsorbate-substrate interaction on metals
organic compounds may undergo structural changes upon
adsorption.\cite{braun_ss01,kang_ss00} In this context we chose to study
perfluorinated copper-phthalocyanine (F$_{16}$CuPc, see
Fig.~\ref{fig:xswsetup}a) on Cu(111) and Ag(111) using the X-ray standing-wave
(XSW) technique.\cite{zegenhagen_ssr93,bedzyk_review,vartanyants_rpp01,%
cheng_prl03,kilian_prb02}
As one of the best air-stable organic n-type semiconductors F$_{16}$CuPc is a
very promising material for future
applications.\cite{forrest_apl01,bao_jacs98} The adsorption of F$_{16}$CuPc,
i.e.\ the bonding distances and possible distortions resulting from the
interaction with the metal electrons, is very relevant as the charge transfer
from and into the metal strongly depends on the structure of the first
molecular layer. 

This paper is organized as follows: In Sec.~\ref{sec:experiment} we describe
the experimental setup and procedures. Sec.~\ref{sec:results} presents our XSW
results on F$_{16}$CuPc with particular emphasis on the data analysis and
non-dipolar contributions.  In Sec.~\ref{sec:discussion} we discuss several
aspects and implications of the results.  Sec.~\ref{sec:conclusion} concludes
this work with a brief summary.

\section{Experimental Details}
\label{sec:experiment}

\subsection{General}

The experiments were carried out at beamline ID32 of the European Synchrotron
Radiation Facility (ESRF) in Grenoble, see Fig.~\ref{fig:xswsetup}b for
details of the experimental setup.  The molecular films of F$_{16}$CuPc were
prepared and studied \textit{in situ} using a multi-purpose ultra-high vacuum
chamber with several analytical components (base pressure $2\times
10^{-10}\,$mbar).

\subsection{Sample preparation}

The Cu(111) and Ag(111) single crystals were mounted on a
variable-temperature, high-precision manipulator.  Repeated cycles of argon
ion bombardment and annealing at $600-700\,$K resulted in clean and largely
defect-free surfaces as has been verified by XPS and LEED measurements.  The
F$_{16}$CuPc material supplied by Aldrich Chemical Co.\ was purified by
gradient sublimation.  Using a thoroughly outgassed Knudsen cell the molecules
were evaporated at typical rates of less than 1~ML/min with the substrate at
$300\,$K. Each evaporation process was controlled with a calibrated quartz
crystal microbalance close to the substrate.

\subsection{Data acquisition}

While the photon energy was scanned through the first order back-reflection
condition for Cu(111) and Ag(111) around 2980 eV and 2630 eV, respectively,
X-ray standing wave signals were recorded.  For this purpose a vertically
mounted hemispherical electron analyzer (Physical Electronics) at an angle of
$45^\circ$ relative to the incoming X-ray beam acquired series of energy
resolved photoemission spectra.


\begin{figure}[htbp]
  \centering 
  \includegraphics[width=8cm]{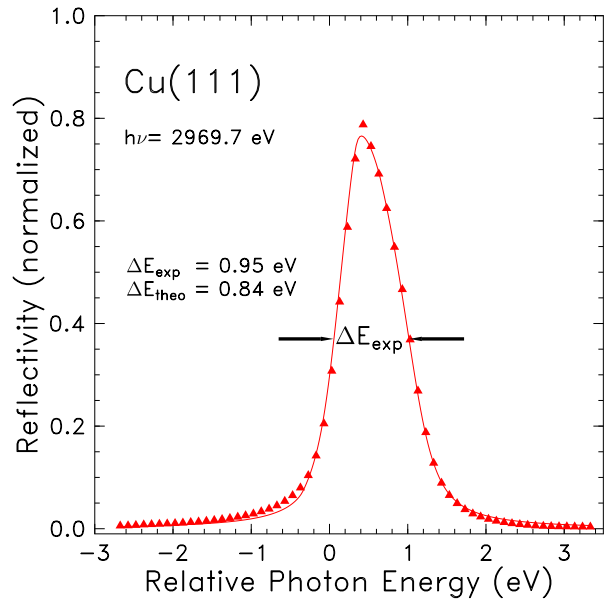}
  \caption{(Color online) Normal incidence reflectivity curve around the [111]
    Bragg reflection of the copper substrate. The solid line represents the
    reflectivity calculated by dynamical diffraction theory with additional
    broadening due to the mosaicity of the sample and the finite monochromator
    resolution. The origin of the relative energy scale used throughout this
    article refers to the Bragg peak position as it would be observed without
    refraction inside the crystal.}
  \label{fig:darwin}
\end{figure}

After positioning the sample the X-ray reflectivity was measured with a
photodiode mounted at a small angle relative to the incoming beam.  As
illustrated for Cu(111) in Fig.~\ref{fig:darwin} we observed the first-order
Bragg peaks whose position and shape can be described very well within the
framework of dynamical diffraction theory.  Since noble metal crystals are
known to exhibit a certain mosaic spread that broadens the Bragg peak, we
always monitored the reflectivity signal to identify a suitable position on
the substrate before doing the XSW experiment.  Given an intrinsic width of
$0.84$\,eV derived from dynamical diffraction theory for a defect free crystal
we regard the observed value of $0.95$\,eV as indication of sufficient crystal
perfection.

\section{Results and Analysis}
\label{sec:results}

\subsection{Photoemission analysis}

\begin{figure*}[htbp]
  \centering 
  \includegraphics[width=14cm]{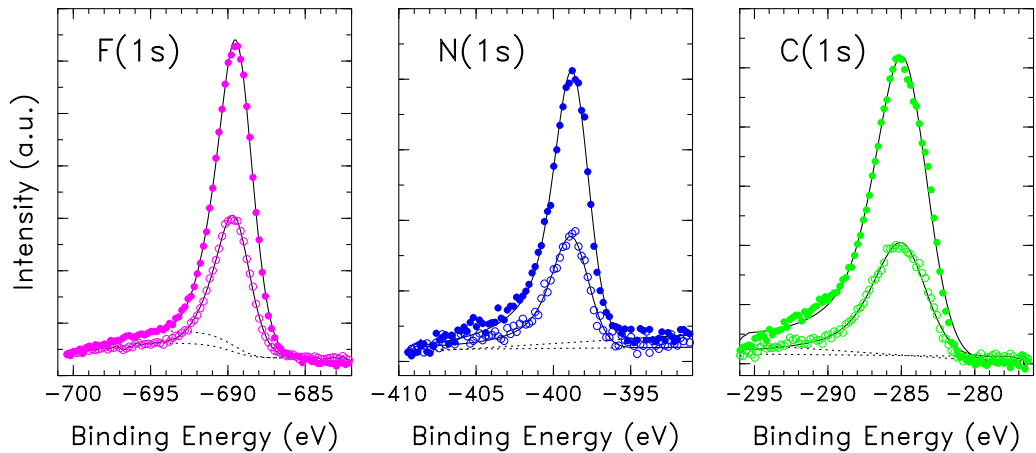}
  \caption{(Color online) Photoemission core-level lines of fluorine,
    nitrogen, and carbon taken on a submonolayer of F$_{16}$CuPc on Cu(111).
    The complete XSW series are analyzed by fitting a Voigt-like asymmetric
    line shape (solid line) and a suitable background (dashed line) to the
    spectra.  Closed symbols refer to a photon energy on the Bragg condition,
    whereas open symbols correspond to an energy $1\,$eV below. With improved
    energy resolution we are able to distinguish two components in the C(1s)
    region corresponding to different chemical environments of the carbon
    atom.}
  \label{fig:xpsspectra}
\end{figure*}

In order to extract the XSW signal a thorough analysis of all photoemission
spectra is required. As shown in Fig.~\ref{fig:xpsspectra} a Voigt-like
asymmetric line shape and an independently scaled Shirley-type background
describe the experimental C(1s), N(1s), and F(1s) core-level spectra very
well.\footnote{The energy level splittings and shake-up states that have been
  observed on F$_{16}$CuPc before\cite{ottaviano_jesrp00} cannot be resolved.}
In particular, we found the careful subtraction of the strongly photon energy
dependent inelastic background essential. By taking integrated peak
intensities and normalizing to the incoming photon flux we obtained the
photoelectron yield datasets which are suitable for the XSW analysis.

Further insight can be gained from spectroscopic observations on the monolayer
system of F$_{16}$CuPc. Importantly, no significant changes in the peak
position or line shapes were observed during the XSW experiment, indicating
that the molecules do not fragment due to radiation damage.  Moreover, the
stoichiometry of the adsorbed molecules can be determined by comparing
relative photoemission intensities. After normalizing the integrated off-Bragg
intensity by the photoionization cross sections, the core-level lines shown in
Fig.~\ref{fig:xpsspectra} give a stoichiometric ratio which corresponds within
the error bars to the F$_{16}$CuPc composition, see Tab.~\ref{tab:xpsresults}.
Similarly, the surface coverage in the monolayer regime was calibrated by
evaluating intensities of a substrate and adsorbate signal.

\begin{table}[htbp]
  \centering
  \begin{ruledtabular}
  \begin{tabular}{l|d|D{.}{.}{4.10}|d}
    & \multicolumn{1}{c|}{$I$ (norm.)} &
    \multicolumn{1}{c|}{$\sigma$ (Mb)\footnote{taken from Ref.~\onlinecite{verner_ant93}}} &
    \multicolumn{1}{c}{$I/\sigma$ (norm.)} \\[1ex]
    \hline
    C(1s) & 100.0  & 1.23\times 10^{-3} & 32.0 \\
    N(1s) &  44.6  & 2.03\times 10^{-3} & 8.6 \\
    F(1s) & 198.4  & 4.96\times 10^{-3} & 15.7 \\
  \end{tabular}
  \end{ruledtabular}
  \caption{Stoichiometry of the adsorbate derived from photoemission
    intensities: When normalizing the raw intensity $I$ obtained from the
    datasets shown in Fig.~\ref{fig:xpsspectra} by the photoionization
    cross section $\sigma$ a composition close to the sum formula
    F$_{16}$C$_{32}$N$_{8}$ is derived.}
  \label{tab:xpsresults}
\end{table}

\subsection{XSW analysis}
\subsubsection{Basic principles}

The variation of the photoelectron yield observed from molecular adsorbates
while scanning the photon energy through the Bragg condition holds structural
information that can be analyzed
quantitatively.\cite{zegenhagen_ssr93,bedzyk_review} However, it has been
shown\cite{jackson_prl00,schreiber_ssl01,nelson_prb02} that depending on the
experimental conditions the dipole approximation of photoemission is not
generally applicable to the analysis of X-ray standing wave data.
Higher-order terms contributing to the photoemission yield must not be
neglected for low-Z elements and typical photon energies of several keV.
Therefore the normalized photoelectron yield $Y_p(\Omega)$ is not simply
proportional to the standing wave intensity, as for the pure dipolar case.
Instead, a generalized relation\cite{vartanyants_ssc00}
\begin{equation}
  Y_p(\Omega ) = 1 + S_R R + 2 |S_I|\sqrt{R} f_H \cos(\nu- 2\pi P_H +\psi)
  \label{eq:xsw_yield}
\end{equation}
that includes first-order corrections has to be used. Here the structural
parameters $f_H$ and $P_H$ are the coherent fraction and position related to
the $\mathit{H^{th}}$ Fourier component of the adsorbate atomic density. The
photon energy dependent reflectivity is described in terms of its absolute
value $R$ and phase $\nu$ between the incoming and outgoing waves.  $S_R$ and
$S_I=|S_I|\exp (i \psi)$ represent the higher-order contributions in the
photoemission matrix element.\cite{vartanyants_ssc00} Therefore they generally
depend on the experimental geometry, the element number, the photon energy,
and orbital symmetry of the initial state. Only within the dipole
approximation with $S_R=1$, $|S_I|=1$, and $\psi=0$ Eq.~(\ref{eq:xsw_yield})
reduces to the more familiar form.\cite{zegenhagen_ssr93,bedzyk_review}

In case of a back-scattering geometry as used throughout our experiments these
three non-dipolar parameters are not independent.\cite{vartanyants_ssc00} Due
to an additional constraint, i.e.\ 
\begin{equation}
   |S_I| = \frac{1}{2}(S_R + 1) \sqrt{1 + \tan^2 \psi},
  \label{eq:SR_relation}
\end{equation}
values for only two non-dipole parameters have to be established to determine
the structural XSW parameters $f_H$ and $P_H$.  With $0\leq P_H \leq 1$ and
$d_0$ as the distance of the substrate Bragg planes we derive the relative
positions $d_H$ of the adsorbate atoms to be $d_H=d_0(1+P_H)$.

\subsubsection{Incoherent films}
\label{sec:results-a}

For thicker films of F$_{16}$CuPc (coverage~$\geq$10~ML) the averaging over
many different positions leads to an effectively incoherent
film\cite{schreiber_ssl01,jackson_prl00}, and with the resulting $f_H=0$
Eq.~(\ref{eq:xsw_yield}) reduces to
\begin{equation}
  Y_p(\Omega ) = 1 + S_R R.
  \label{eq:xsw_yield_incoh}
\end{equation}

\begin{figure}[htbp]
  \centering 
  \includegraphics[width=8cm]{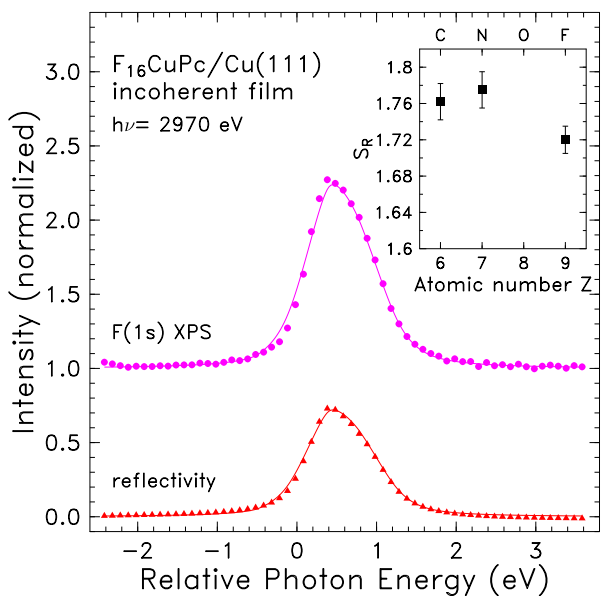}
  \caption{(Color online) Typical X-ray standing wave scan on an incoherent
    film of F$_{16}$CuPc on Cu(111). By using Eq.~(\ref{eq:xsw_yield_incoh})
    the non-dipole parameter $S_R$ can be determined from the experimental XSW
    signal. The small deviations of the XSW fit from the experimental data can
    be traced back to several parameters which affect the broading of the
    reflectivity and photoelectron signal. In particular, the different
    (angular and spatial) resolution functions in these measurements feature a
    slightly broader Bragg peak. The inset shows results for the corresponding
    core-levels of the different elements in F$_{16}$CuPc with realistic error
    bars.}
  \label{fig:cu111incoh}
\end{figure}

As has been demonstrated before\cite{lee_ss01,schreiber_ssl01} the non-dipole
parameter $S_R$ can be determined by measuring the reflectivity and the XSW
yield of the different atomic species. The relatively strong photoemission
signals observed from multilayers of F$_{16}$CuPc provide datasets with almost
negligible statistical noise that can be analyzed according to
Eq.~(\ref{eq:xsw_yield_incoh}).\footnote{Indeed, $S_R$ depends on the
  experimental geometry and in particular the electron emission angle. The
  finite angular resolution of the hemispherical electron analyzer is
  neglected here.} On the basis of fits as the one shown in
Fig.~\ref{fig:cu111incoh} we obtain $S_R$-results on C(1s), N(1s), and F(1s)
for first-order back-reflection energies of Cu(111) and Ag(111), see
Fig.~\ref{fig:cu111incoh} and Tab.~\ref{tab:nondip}. Our data are in good
agreement with previous experimental results on Cu(111)\cite{lee_ss01,
  jackson_prl00} and ab-initio calculations\cite{bechler_pra89,nefedov_dp99}.
Given the experimental results, i.e.\ \mbox{$1.59 \leq S_R \leq 1.77$} for the
different elements, the non-dipolar enhancement of the photoelectron yield is
a key factor for the structural XSW analysis.

\subsubsection{Coherent films}
\label{sec:results-b}

The photoelectron yield observed from a monolayer of F$_{16}$CuPc molecules is
directly related to the the spatial phase of the XSW field at the atomic
positions. Thus with $f_H>0$ the coherent positions $P_H$ can be determined,
provided that the non-dipolar terms $|S_I|$ and $\psi$ are taken into account.
By introducing effective quantities
\begin{equation}
  f_\mathit{eff} = |S_I| f_H \qquad \textnormal{and} \qquad
  P_\mathit{eff} = P_H - \psi / 2\pi 
  \label{eq:effec_par} 
\end{equation}
in Eq.~(\ref{eq:xsw_yield}) the photoemission yield may be written as
\begin{equation}
  Y_p(\Omega ) = 1 + S_R R + 2 \sqrt{R} f_\mathit{eff}
   \cos (\nu - 2\pi P_\mathit{eff}).  
  \label{eq:effec_XSWyield}
\end{equation}
Using the previously measured $S_R$-values the effective parameters defined in
Eqs.~(\ref{eq:effec_par}) can now be derived directly from experimental
photoelectron yield data. Therefore Eq.~(\ref{eq:effec_XSWyield}) has been the
`working equation' for analyzing the XSW data.

\begin{figure}[htbp]
  \centering 
  \includegraphics[width=8cm]{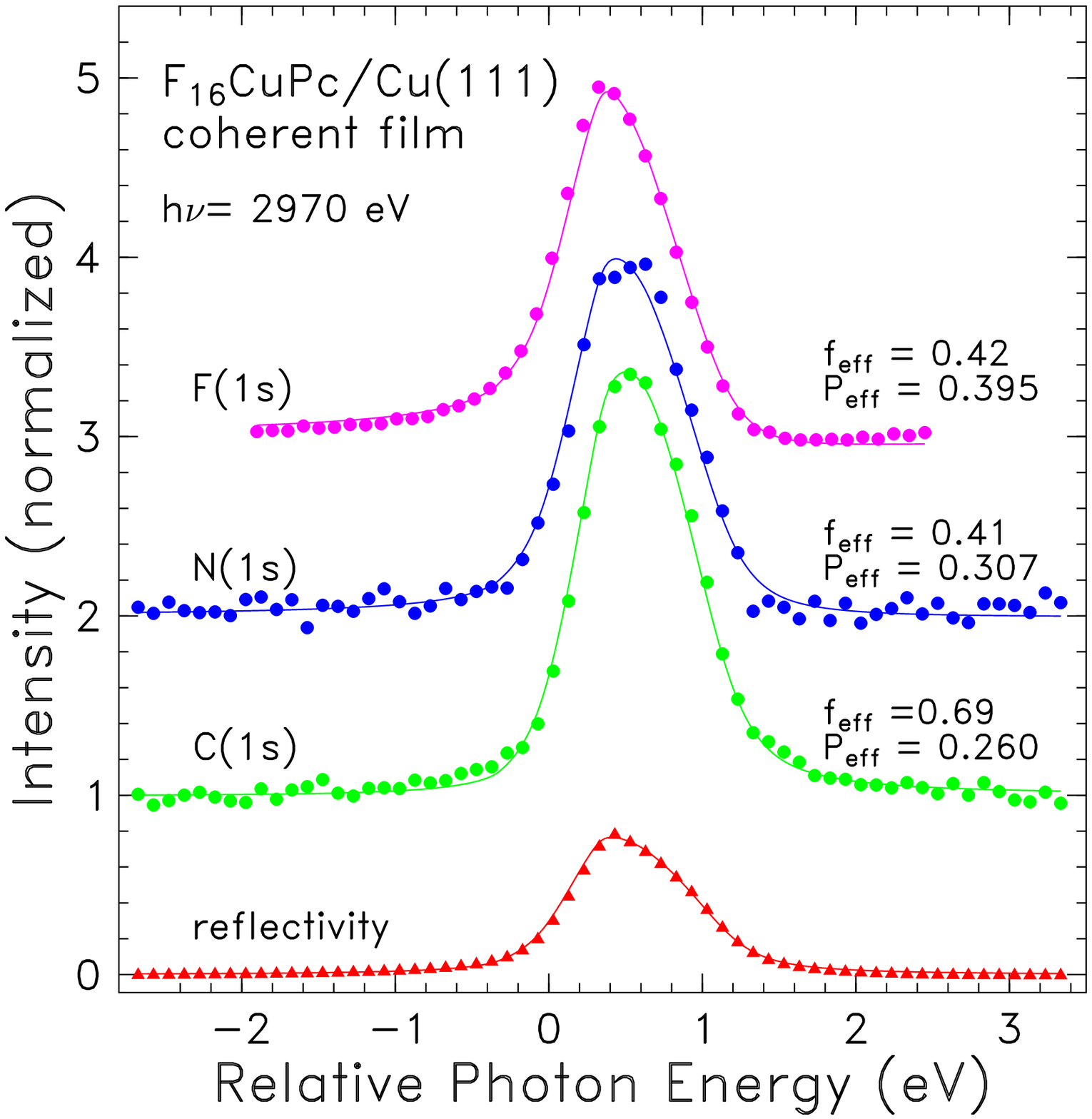}
  \vfill
  \includegraphics[width=8cm]{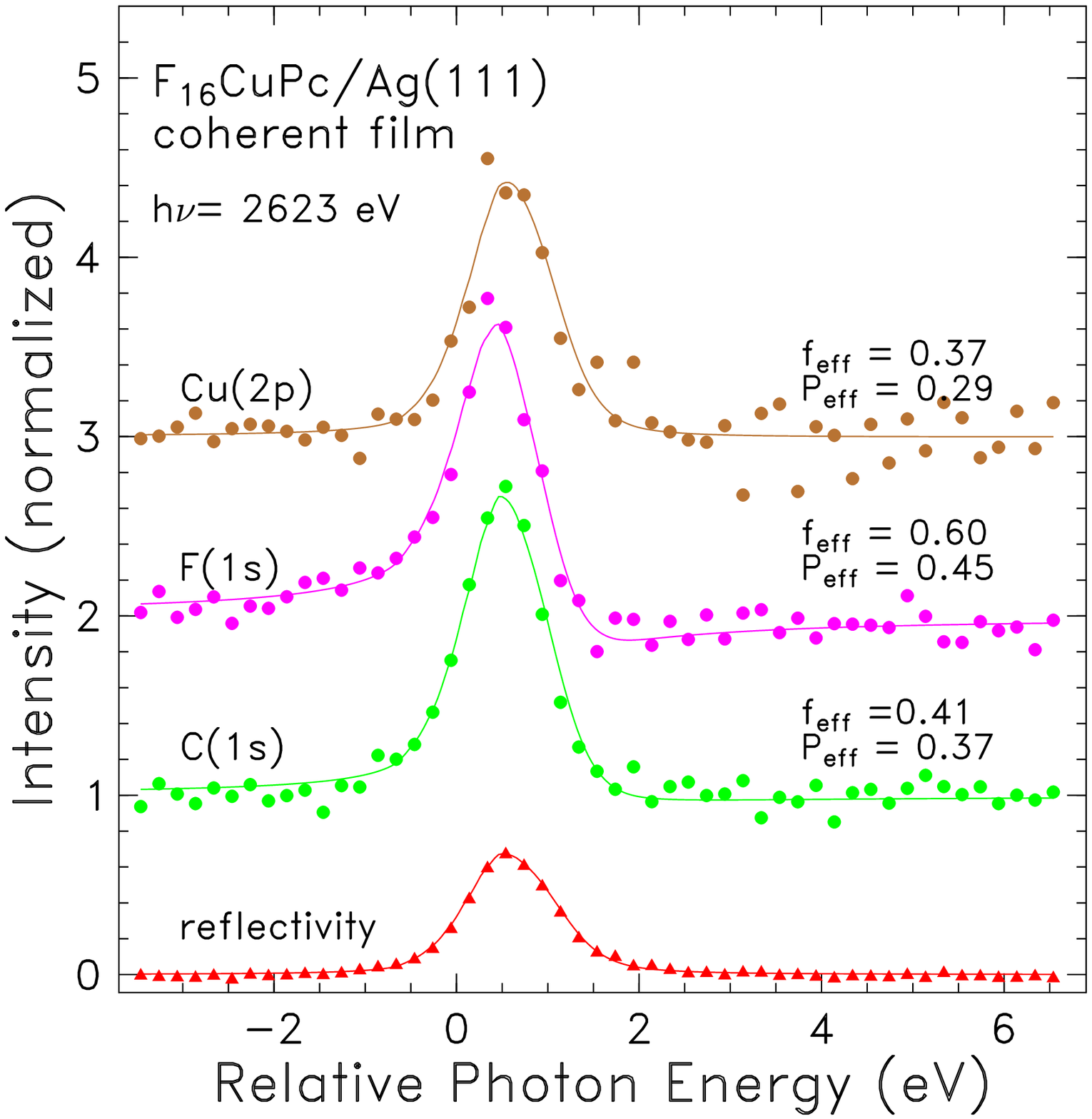}
  \caption{(Color online) X-ray standing wave scans on a submonolayer of
    F$_{16}$CuPc on Cu(111) and Ag(111). The effective coherent fraction
    $f_\mathit{eff}$ and coherent position $P_\mathit{eff}$ are determined by
    fitting Eq.~(\ref{eq:effec_XSWyield}) to the experimental data. For
    clarity the datasets for N(1s), F(1s), and Cu(2p) are plotted with an
    offset.}
  \label{fig:cu111coh}
\end{figure}

The XSW scans on F(1s), N(1s), C(1s) presented in
Fig.~\ref{fig:cu111coh}~(top) were taken on a submonolayer of F$_{16}$CuPc on
Cu(111).  As a first, more qualitative result we note the similar overall
shape of these XSW scans which indicate comparable coherent positions and thus
a lying down configuration of the molecules. The low noise level achieved in
these measurements, however, allows us to resolve small, but significant
differences in the shape of the XSW signals: Compared to the carbon or
nitrogen signal the fluorine yield shown in Fig.~\ref{fig:cu111coh}~(top)
exhibits a more pronounced tail on the low-energy side.  Accordingly,
different coherent contributions are found by least-square fits on the basis
of Eq.~(\ref{eq:effec_XSWyield}) which yield a coherent position of
$P_\mathit{eff}=0.395$ for fluorine and $P_\mathit{eff}=0.260$ for carbon.

Likewise we obtained X-ray standing-wave signals from a coherent layer of
F$_{16}$CuPc on Ag(111). The XSW scans on C(1s), F(1s), and Cu(2p$_{3/2}$)
shown in Fig.~\ref{fig:cu111coh}~(bottom) again reveal a lying down
configuration of the molecules. Despite slightly worse statistics in these
data our analysis works well and the fit parameters $f_\mathit{eff}$ and
$P_\mathit{eff}$ can be determined precisely. As on Cu(111) we derive a
markedly larger coherent position $P_\mathit{eff}=0.45$ for fluorine compared
to $P_\mathit{eff}=0.37$ for carbon.  Further details on the resulting
effective parameters both on Cu(111) and Ag(111) can also be found in
Tab.~\ref{tab:xswresults}.  The exact atomic positions $d_H$, however, cannot
be derived unless the non-dipolar contributions are separated out.

\subsubsection{Non-dipolar corrections}

In order to retrieve the coherent position $P_H$ and the coherent fraction
$f_H$ from the effective parameters either the additional phase $\psi$ or
$|S_I|$ has to be known. Importantly, in case of initial $s$-state symmetry
this problem can be overcome because $\psi$ is directly related to the partial
phase shift $\Delta = \delta_d - \delta_p$ between the possible final $p$- and
$d$-states of the photoexcitation process.  Since it can be shown that
\begin{equation}
  \tan \psi = \frac{S_R-1}{S_R+1} \: \tan \Delta ,
  \label{eq:delta}
\end{equation}
the XSW phase $\psi$ is a simple and unique function of the partial phase
shift $\Delta$. Using an averaged experimental value of $S_R \approx 1.75$ as
a first estimate we hence find $\psi \approx 0.27 \Delta$.

For each element and electron energy phase shifts $\Delta$ are determined
independently by means of relativistic \textit{ab initio}
calculations.\cite{jablonski_nist02} Our results as given in
Tab.~\ref{tab:nondip} are in excellent agreement with previous theoretical
efforts (Ref.~\onlinecite{bechler_pra89} and Fig.~7 in
Ref.~\onlinecite{lee_ss01}).  The corresponding non-dipolar XSW phases $\psi$
for carbon, nitrogen, and fluorine turn out to be relatively small and
similar, with only minor impact on the effective coherent
positions.\footnote{We note that the Auger signal of a coherent layer cannot
  easily be used to determine $\psi$ experimentally.  For several reasons it
  is difficult to extract the pure dipolar signal which would be needed for
  comparison: First, coherent Auger excitations by photoelectrons coming from
  the substrate are generally not negligible.  Second, the weak and broad
  Auger lines require very long integration times and a thorough intensity
  analysis to be useful. With a typical signal-to-background ratio of only
  0.05 for the fluorine peaks even minute errors in the definition of the
  background level or shape result in significant problems.  This strongly
  photon energy dependent background of mostly inelastically scattered
  substrate electrons has to be subtracted consistently for each spectrum in
  the XSW series.  Third, as it turns out that the phase difference $\psi$
  between Auger and XPS signal is relatively small, the determination of
  $\psi$ means subtracting similar numbers whereby the experimental errors
  propagate very unfavorably.}  Therefore we find $|S_I|\approx \frac{1}{2}
(S_R+1)$ as a good approximation to Eq.~(\ref{eq:SR_relation}) with $S_R$ and
$|S_I|$ being the truly important non-dipolar parameters in our experiment.

\begin{table}[htbp]
  \centering
  \begin{ruledtabular}
    \begin{tabular}{l|D{.}{.}{1.6}|D{.}{.}{1.6}|D{.}{.}{1.6}|D{.}{.}{1.6}|D{.}{.}{1.6}}
      & \multicolumn{3}{c|}{Cu(111)} & \multicolumn{2}{c}{Ag(111)} \\[1ex] 
      & \multicolumn{1}{c|}{C(1s)} & \multicolumn{1}{c|}{N(1s)} &
      \multicolumn{1}{c|}{F(1s)} & \multicolumn{1}{c|}{C(1s)} &
      \multicolumn{1}{c}{F(1s)} \\[1ex]
      \hline
      $S_R$    & 1.76(1) & 1.77(1)  & 1.72(1) & 1.74(1)  & 1.59(1)  \\[1ex] 
      $\Delta$ & -0.199 & -0.236 & -0.321 & -0.211 & -0.346 \\
      $\Delta$\footnote{taken from Ref.~\onlinecite{lee_ss01}} & -0.20 & -0.24 &
      -0.33 &  &  \\
      $\psi$ & -0.055 & -0.067 & -0.088 & -0.058 & -0.082 \\
      $|S_I|$ & 1.382 & 1.388 & 1.365 & 1.372 & 1.299 \\
    \end{tabular}
  \end{ruledtabular}
  \caption{Non-dipolar parameters: The $S_R$-values are derived
    experimentally from incoherent films, whereas $\Delta$ is obtained from
    ab-initio calculations\cite{jablonski_nist02}. For comparison
    values taken from Ref.~\onlinecite{lee_ss01} are given. Evaluation of
    Eq.~(\ref{eq:delta}) and Eq.~(\ref{eq:SR_relation}) then gives $\psi$ and
    $|S_I|$, respectively.}
  \label{tab:nondip}
\vspace{3ex}
  \begin{ruledtabular}
    \begin{tabular}{l|D{.}{.}{1.6}|D{.}{.}{1.6}|D{.}{.}{1.6}|D{.}{.}{1.6}|D{.}{.}{1.6}}
      & \multicolumn{3}{c|}{Cu(111)} & \multicolumn{2}{c}{Ag(111)} \\[1ex] 
      & \multicolumn{1}{c|}{C(1s)} & \multicolumn{1}{c|}{N(1s)} &
      \multicolumn{1}{c|}{F(1s)} & \multicolumn{1}{c|}{C(1s)} &
      \multicolumn{1}{c}{F(1s)} \\[1ex]
      \hline
      $f_\mathit{eff}$ & 0.69(4) & 0.41(4) & 0.42(3) & 0.41(6) & 0.60(4) \\
      $P_\mathit{eff}$ & 0.260(5)  & 0.308(8)  & 0.395(9) & 0.370(19) & 0.450(12) \\
      $f_H$            & 0.50(1)   & 0.30(1)   & 0.31(1)  & 0.30 & 0.46  \\
      $P_H$            & 0.251(5)  & 0.297(8)  & 0.381(9) & 0.380 & 0.463
      \\[1ex] 
      $d_H$            & 2.61\,\textnormal{\AA} & 2.70\,\textnormal{\AA} &
      2.88\,\textnormal{\AA}  & 3.25\,\textnormal{\AA} & 3.45\,\textnormal{\AA}
    \end{tabular}
  \end{ruledtabular}
  \caption{XSW results taken on a submonolayer of F$_{16}$CuPc on Cu(111)
      and Ag(111): By taking into account the non-dipolar effects we derive the
      atomic  position $d_H$ relative to the Bragg planes of the substrate. In
      parentheses we give the statistical uncertainties of the
      parameters. With systematic uncertainties included we estimate the error
      bar of $d_H$ to be $\pm 0.07\,$\AA{} on copper and $\pm 0.10\,$\AA{} on
      silver.}
  \label{tab:xswresults}
\end{table}

Finally, we are now able to deduce the coherent fractions $f_H$ and coherent
positions $P_H$ which yield the adsorbate bonding distances $d_H$ relative to
the Bragg planes of the substrate. On Cu(111) we find $d_H = 2.61\,$\AA{} for
carbon, whereas the fluorine atoms reside at $d_H = 2.88\,$\AA, i.e.\ 
$0.27\,$\AA{} above the central benzene rings of the F$_{16}$CuPc molecule.
With $d_H=2.70\,$\AA{} we locate nitrogen in an intermediate position somewhat
closer to the carbons.  The coherent fractions we derive on copper are nearly
identical for fluorine and nitrogen, yet larger for fluorine.  On Ag(111) we
obtain $d_H = 3.25\,$\AA{} for carbon, and $d_H = 3.45\,$\AA{} for fluorine.
Again this difference of $0.20\,$\AA{} between both elements reveals a
noticeable distortion of F$_{16}$CuPc with the fluorine atoms above the plane
defined by the inner carbon rings.

\subsubsection{Error analysis}

\begin{figure}[htbp]
  \centering 
  \includegraphics[width=8cm]{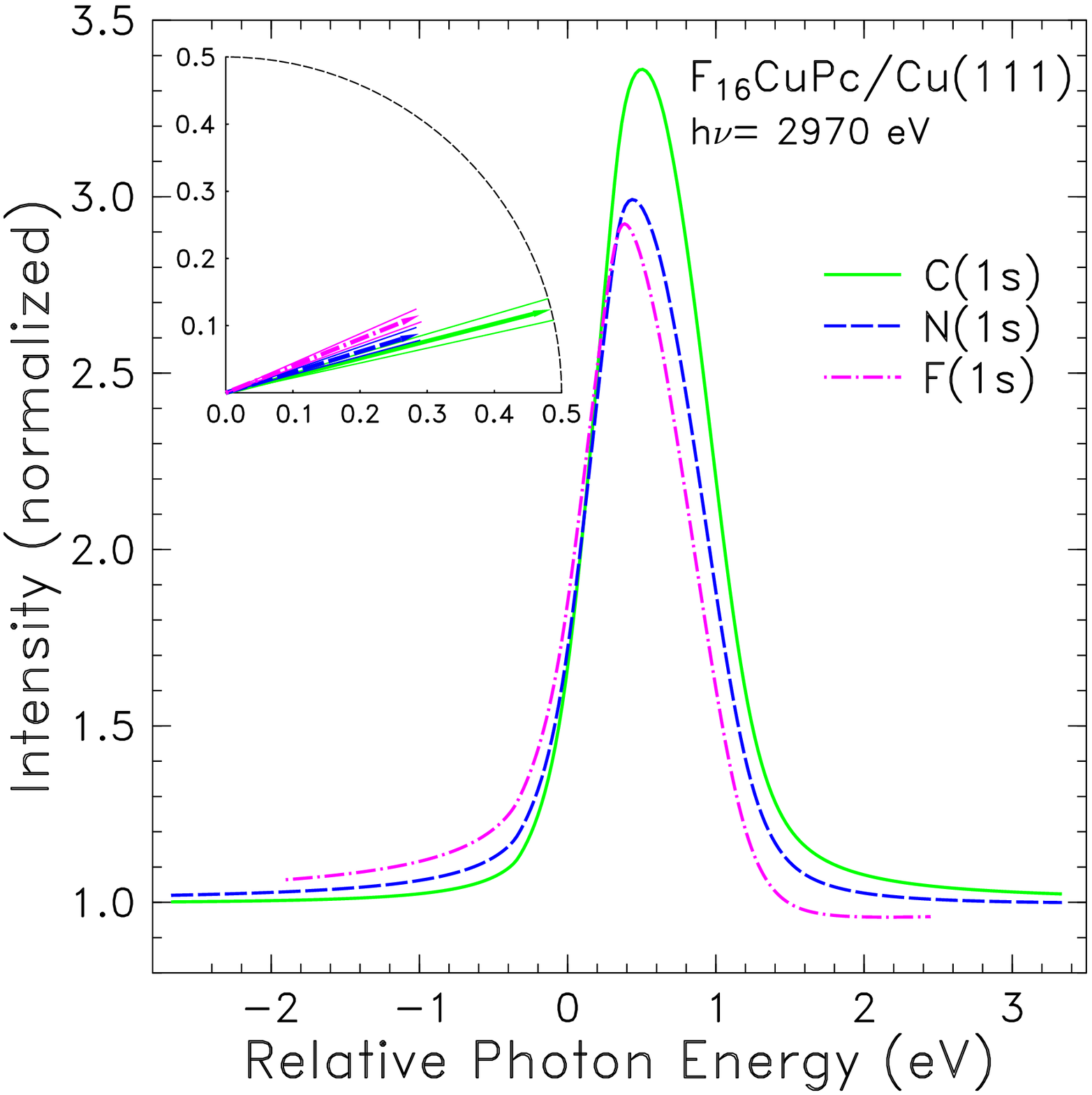}
  \includegraphics[width=8cm]{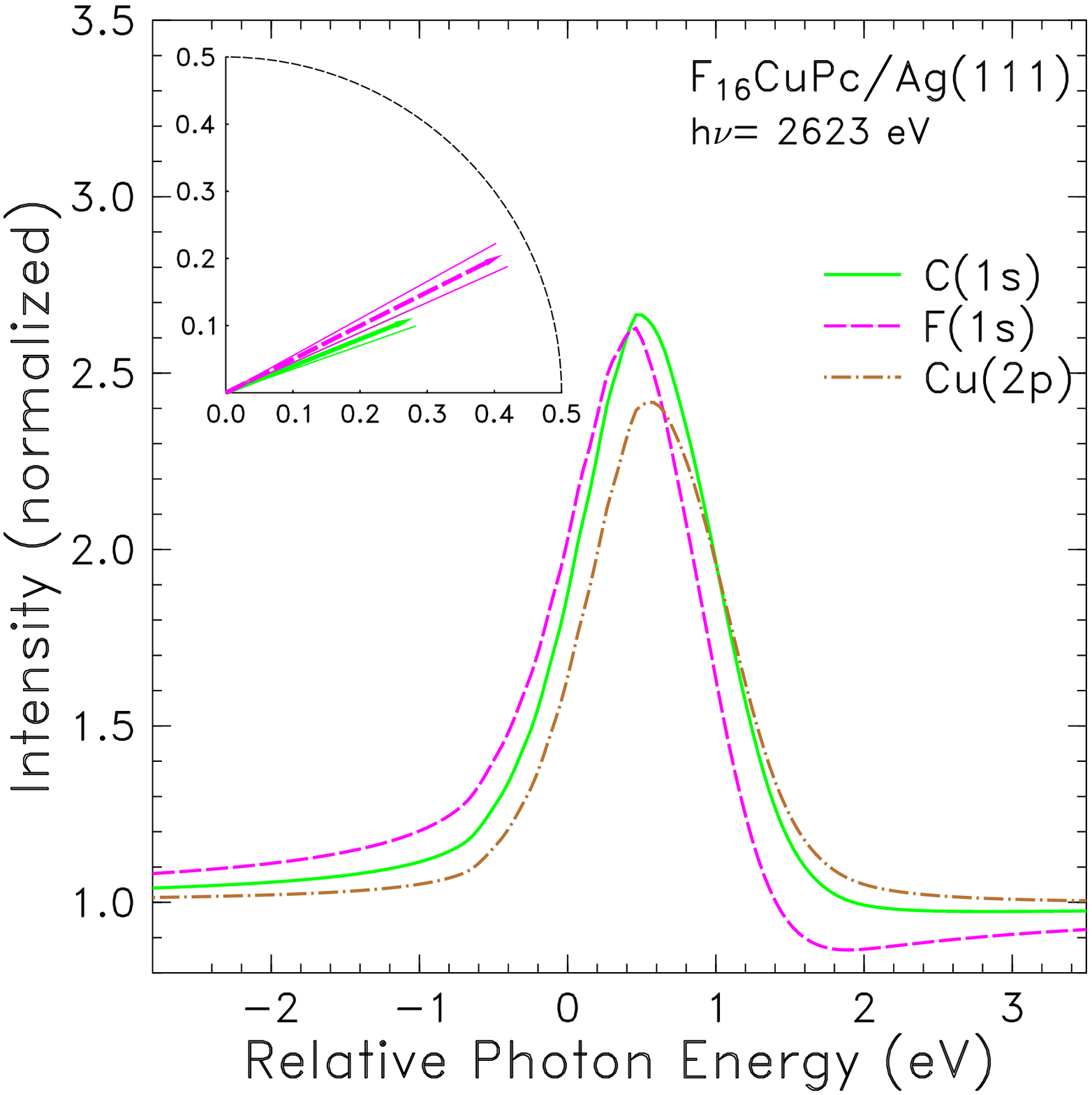}
  \caption{(Color online) Comparison of XSW fits of F$_{16}$CuPc on Cu(111)
    and Ag(111) for C(1s), N(1s), and F(1s) with different tails on the low
    and high energy side of the XSW signal. The inset shows the corrected
    values for $f_H$ and $P_H$ with realistic error bars in the Argand diagram
    corresponding to $\Delta d_H=\pm 0.07\,$\AA{} ($\pm 0.10\,$\AA) on copper
    (silver).}
  \label{fig:cu111coh_corr}
\end{figure}

Showing the relevant fits to our XSW data on copper and silver
Fig.~\ref{fig:cu111coh_corr} demonstrates the obvious differences between
these datasets. In order to assess our XSW results and decide whether the
different bonding distances are significant a careful error analysis is
necessary.
We included Poisson-like errors as weighting factors in the fitting procedure
of Eq.~(\ref{eq:effec_XSWyield}).  As shown in Tab.~\ref{tab:xswresults} the
obtained error bars for the coherent position $P_\mathit{eff}$ are usually
quite small. The corresponding uncertainties in the adsorbate positions $d_H$
therefore amount to barely $\pm 0.01 \ldots 0.02\,$\AA{} for datasets as those
shown in Fig.~\ref{fig:cu111coh}.

Systematic errors of different origin, however, are much more difficult to
quantify. Experimental insufficiencies and incorrect data analysis practices
may inflict deviations from the 'true' XSW signal. Because of the fixed focus
of the electron analyzer, for example, a drifting X-ray beam on the sample can
be precarious.  Similarly, a wrong decomposition of the photoemission spectra
causing erroneous XSW intensities can be misleading.  Nevertheless, the
pronounced tail on the low-energy side of the fluorine XSW signal as seen in
Fig.~\ref{fig:cu111coh} is consistently observed from monolayers F$_{16}$CuPc
on Cu(111) and Ag(111).  Based on our experience with many different datasets
we consider the systematic error of $d_H$ to be dominant resulting in an
accuracy of typically $\pm 0.05 \ldots 0.10\,$\AA.  We therefore conclude that
the elevated positions of the fluorine atoms relative to the central benzene
rings and the nitrogen atoms are significantly beyond the combined error bars.

\section{Discussion of Results}
\label{sec:discussion}

Like many other molecules with extended $\pi$-electron systems F$_{16}$CuPc
adsorbs in a lying down configuration on Cu(111) and Ag(111) forming a rather
stable adsorbate complex.  This behavior might be explained by the formation
of interface states derived from the delocalized $\pi$-electrons in
F$_{16}$CuPc.\cite{evangelista_ss04} By aiming at a large orbital overlap with
the electron cloud of the substrate, the molecules naturally adopt the lying
down configuration as the energetically most favorable position.
As we observe XSW signals with coherent fractions $0.3\leq f_H \leq 0.5$, the
corresponding disorder within the adlayer is significant. Given the size and
symmetry of F$_{16}$CuPc, this appears to be the result of a statistical
misalignment rather than a uniform tilt of all molecules.\footnote{Already a
  tilt angle of 1$^\circ$ would no longer be consistent with the fluorine XSW
  data shown in Fig.~\ref{fig:cu111coh}.}  Since the lateral structure of
F$_{16}$CuPc might be neither simple nor entirely static\cite{grand_ss96}, the
atomic positions reported here are element and time averaged results. Hence we
regard the positional spread within the atomic ensembles to be instrinsic to
the complex structure of the adsorbate.

The exact bonding distances of F$_{16}$CuPc, to our knowledge determined for
the first time here, are more difficult to interpret. As a first attempt one
might compare our results with the van der Waals radii $r_\mathit{vdW}$ of the
different atoms, given in Tab.~\ref{tab:radii}.  These values, derived from
contact distances between non-bonding atoms do not take chemical bonding or
charge redistribution into account.  In fact, in compounds of different atoms
the radius strongly depends on the chemical bonding. In particular due to the
presence of fluorine, the most electronegative element, one has to expect
significant deviations from these numbers. Not too surprisingly, therefore,
the bonding distances do not agree with added values of $r_\mathit{vdW}$. More
instructive, however, is a comparison with experimental data available for
similar systems. The simplest and probably best studied aromatic adsorbate
system is benzene.  On the transition metal surfaces Ni(111) and Ru(0001)
generally smaller values for the carbon positions are found, i.e.\ 
$1.81\,$\AA{} on nickel\cite{kang_ss00} and $2.11\,$\AA{} on
ruthenium\cite{braun_ss01}.  Examples of more complex molecules are
PTCDA\cite{krause_jcp03} with a bonding distance of $2.85 \pm 0.05\,$\AA{} and
NTCDA\cite{stanzel_ssl04} with $3.02\pm 0.02\,$\AA{} both on Ag(111), i.e.\ 
values comparable to our results.

\begin{table}[htbp]
  \centering
  \begin{ruledtabular}
    \begin{tabular}{l|D{.}{.}{3.4}D{.}{.}{3.4}D{.}{.}{3.4}D{.}{.}{3.4}D{.}{.}{3.4}D{.}{.}{3.4}}
    & \multicolumn{1}{c}{C} & \multicolumn{1}{c}{N} & \multicolumn{1}{c}{F} &
    \multicolumn{1}{c}{Cu} & \multicolumn{1}{c}{Ag} \\ 
    \hline
    $r_\mathit{atomic}$ (\AA) & 0.70 & 0.65 & 0.50 & 1.35 & 1.60 \\
    $r_\mathit{vdW}$ (\AA)    & 1.70 & 1.55 & 1.47 & 1.40 & 1.72 \\
  \end{tabular}
  \end{ruledtabular}
  \caption{Atomic and van der Waals radii of the relevant atoms in
    F$_{16}$CuPc. These van der Waals radii $r_\mathit{vdW}$ are established
    from contact distances between non-bonding atoms and neglect the molecular
    structure of F$_{16}$CuPc.} 
  \label{tab:radii}
\end{table}

\begin{figure}[htbp]
  \centering 
  \includegraphics[height=3.5cm]{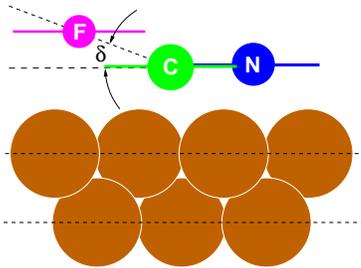}
  \caption{(Color online) Illustration of the F$_{16}$CuPc adsorption geometry
    on Cu(111) (not to scale): Here the fluorine atoms reside $0.27\,$\AA{}
    above the benzene rings forming an average angle of
    $90+\delta=101.5^\circ$ relative to the axis connecting the outer carbons
    with the substrate.}
  \label{fig:f16cupcbend}
\end{figure}

The different atomic positions within the molecule cannot fully be explained
by means of a simple model which does not take the molecular structure of
F$_{16}$CuPc and the presence of the substrate adequately into account. As
discussed in a recent theoretical work\cite{favot_epl00}, however, the
distortion might be related to a partial rehybridization of the carbon atoms
as they change from the $sp^2$-hybridization in the free molecule towards a
more tetrahedral $sp^3$-symmetry upon adsorption.  A convenient way to
illustrate this concept is to consider the average angle $\delta$ between the
C--F bond and the surface, see Fig.~\ref{fig:f16cupcbend}. Using a C--F bond
length of $1.35\,$\AA{} and our XSW results we can derive an angle of
$90+\delta=101.5\pm4.4^\circ$ for Cu(111) and $90+\delta=98.5\pm6.0^\circ$ for
Ag(111). Both values are considerably closer to the tetrahedral angle of
$109.5^\circ$ that would correspond to a full $sp^3$-symmetry.
%
%
However, theoretical work is required to verify whether the surface
interactions are large enough to promote the adsorbing molecule into a
partially $sp^3$-hybridized state.

Further experiments using different ligands as 'spacers' (e.g.\ replacing F
with Cl, Br, or I) could test this hypothesis and reveal how the interaction
between the central ring structure with the metallic electron cloud is
mediated. We note that a distorted adsorption geometry of F$_{16}$CuPc has
interesting and possibly important implications. Due to the high electron
affinity of fluorine a permanent molecular dipole moment perpendicular to the
substrate surface is created.  This, however, results in an additional
attractive force between the molecules and the metal as the induced image
dipole stabilizes this configuration.

First-principle calculations of the adsorption of F$_{16}$CuPc could also shed
more light on this phenomenon as they would include all important aspects of
these system as e.g.\ the character of the chemical bonding in the molecule,
the partially filled $d$-bands in noble metals, and the central copper atom in
F$_{16}$CuPc. The molecular distortion could then be compared to theoretical
results.  These investigations would not only contribute to a better
understanding of these adsorbate systems, but also provide new clues for areas
such as organic electronics, where the binding of the first molecular layer to
a metal contact strongly influences the interface dipole and the charge
carrier injection.

\section{Summary and Conclusions}
\label{sec:conclusion}

In this study we show that large $\pi$-conjugated F$_{16}$CuPc molecules
adsorb in a lying down, but non-planar configuration on the noble metal
surfaces Cu(111) and Ag(111). A detailed, element-specific analysis of our XSW
data reveals a significant relaxation of the molecules upon adsorption. The
coherent positions $P_H$ of the fluorine and carbon atoms differ beyond the
experimental uncertainties: On copper (silver) the central carbon rings are
located at $d_H=2.61\,$\AA{} ($d_H=3.25\,$\AA) above the substrate, whereas
the outer fluorine atoms are found at $d_H=2.88\,$\AA{} ($d_H=3.45\,$\AA).

We hope that our results will stimulate further experimental and theoretical
work in this area.  Calculations on the adsorbate structure of large molecules
would greatly promote our understanding of these systems and could also
provide new insight in the electronic properties of the organic-inorganic
interface.

\section*{Acknowledgments}

The authors thank the ESRF for providing excellent facilities, N.\ Karl and
J.\ Pflaum for purifying the F$_{16}$CuPc material, and the referee for giving
very useful comments.  This work was financially supported by the EPSRC
(UK).

\bibliography{xsw_cupc}

\end{document}